\documentclass[fleqn,usenatbib]{mnras}

\usepackage{newtxtext,newtxmath}
\usepackage[T1]{fontenc}
\usepackage{ae,aecompl}

\usepackage{graphicx}	% Including figure files
\usepackage{amsmath}	% Advanced maths commands
\usepackage{amssymb}	% Extra maths symbols
\usepackage{CJKutf8}

\title[Constraining outflow geometries]{Constraining the geometry of AGN outflows with reflection spectroscopy}

\author[M. L. Parker et al.]{M. L. Parker,$^{1}$\thanks{E-mail: mparker@sciops.esa.int}
D. J. K. Buisson,$^{2}$
J. Jiang \begin{CJK*}{UTF8}{gbsn}(姜嘉陈)\end{CJK*},$^2$
L. C. Gallo,$^3$
E. Kara,$^4$
\newauthor
G. A. Matzeu$^{1}$ and
D. J. Walton$^{2}$
\\
$^{1}$European Space Agency (ESA), European Space Astronomy Centre (ESAC), E-28691 Villanueva de la Ca\~{n}ada, Madrid, Spain\\
$^{2}$Institute of Astronomy, University of Cambridge, Madingley Road, CB3 0HA Cambridge, UK\\
$^3$Saint Mary’s University, Department of Astronomy \& Physics, 923 Robie Street, Halifax, Canada, B3H 3C3\\
$^4$Department of Astronomy, University of Maryland, College Park, MD 20742, USA
}

\date{Accepted XXX. Received YYY; in original form ZZZ}

\pubyear{2015}

\begin{document}
\label{firstpage}
\pagerange{\pageref{firstpage}--\pageref{lastpage}}
\maketitle

\begin{abstract}
We collate active galactic nuclei (AGN) with reported detections of both relativistic reflection and ultra-fast outflows. By comparing the inclination of the inner disc from reflection with the line-of-sight velocity of the outflow, we show that it is possible to meaningfully constrain the geometry of the absorbing material. We find a clear relation between the velocity and inclination, and demonstrate that it can potentially be explained either by simple wind geometries or by absorption from the disc surface.
Due to systematic errors and a shortage of high-quality simultaneous measurements our conclusions are tentative, but this study represents a proof-of-concept that has great potential.
\end{abstract}


%%%%%%%%%%%%%%%%%%%%%%%%%%%%%%%%%%%%%%%%%%%%%%%%%%

%%%%%%%%%%%%%%%%% BODY OF PAPER %%%%%%%%%%%%%%%%%%

\section{Introduction}

There are two main forms of relativistic spectroscopy in X-ray studies of AGN: relativistic reflection, and ultra-fast outflows. Both rely on detecting red and blue-shifted elemental emission or absorption lines, usually from iron, and both reach velocities of a significant fraction of $c$. These two techniques are rarely used together, so the opportunities afforded by combining the two remain largely unexplored.

Relativistic reflection occurs when relatively cool, dense material in the accretion disc around a black hole is illuminated by X-rays. A featureless X-ray continuum is reprocessed into a series of fluorescent emission lines, absorption edges, and a Compton scattered hump \citep{George91}. This characteristic reflection spectrum is then blurred and shifted by a combination of Doppler shifting, special relativistic boosting, and gravitational redshift \citep{Fabian89}. By measuring the extent of these relativistic effects on the profile of the Fe~K$\alpha$ line \citep[or other lines, e.g.][]{Fabian09, Madej14}, we can infer physical parameters of the black hole and accretion disc, such as the inclination of the disc and the black hole spin parameter \citep[see review by][]{Reynolds14}. 

Ultra-fast outflows (UFOs) are identified by strongly blueshifted absorption lines in X-ray spectra \citep{Chartas02,Pounds03}. They are generally thought to be due to winds from the AGN disc, accelerated by magnetic or radiation pressure \citep{Reeves03, Fukumura15}, and are one possibility for driving AGN feedback \citep[e.g.][]{Fabian12_feedback}. An alternative possibility for producing these features is in an absorbing layer on the surface of the disc \citep{Gallo11_ufos, Gallo13_ufos}, where the extreme velocity arises from the orbital motion of the gas and the absorption is imprinted on the reflection spectrum.
Because we only have one line of sight (LOS) through the gas, it is extremely hard to directly measure the density and hence location of the absorbing material, so we must use other approaches to try and constrain the geometry of the gas.

While there is no reason to expect viewing angle to be the sole determiner of the line of sight velocity, it should certainly have an impact, which depends on the launching mechanism. Radiation pressure driven disc winds are expected to have an equatorial geometry \citep[e.g.][]{Proga00}, so velocity would generally increase with inclination, whereas MHD wind simulations predict higher velocities at low inclinations \citep{Fukumura10}. 
In this work, we examine the possibility of constraining the geometry of UFOs using their inclination dependence, taking the inclination values from reflection modeling.

\section{Sample}
\label{sec_sample}
We performed a literature search to identify sources which have both absorption from an outflow with a well constrained velocity and a well constrained inclination from relativistic reflection spectroscopy. In a few cases \citep[][Parker et al., submitted]{Tombesi11_3c111, Parker17_nature} both are measured simultaneously, but for the majority these measurements are taken from different papers. In general, we prioritise results from papers presenting spectral fitting of an individual source over those where a large sample of sources are analysed. We also prefer more recent papers, as these are likely to have higher quality datasets available, as well as the latest models. The selection of UFO and reflection results is presented in detail below, and the values are given in Table~\ref{tab_data}.
The data and code used for the analysis in Section~\ref{sec_results} are available here:
\url{https://github.com/M-L-Parker/UFO_inclinations}.

\subsection{UFOs}
\label{sec_sample_ufos}
%The majority of UFO detections come from the \citet{Tombesi10} and \citet{Gofford13} sample papers, using \xmm\ and \suzaku , respectively.
%When available, we take prioritize measurements with a physical model such as \textsc{xstar} \citep{Kallman01} over those from fitting Gaussian lines.

To avoid contamination from warm absorbers, we implement a velocity cut-off at 0.033$c$ \citep[10,000~km~s$^{-1}$;][]{Tombesi10}.%, below which we do not include any sources in our sample. 
These may or may not be related to UFOs \citep[see discussion in e.g.][]{Tombesi13_unification, Pounds13, Laha16}, and in general have small velocities or are consistent with the source rest-frame \citep{Laha14}.

The treatment of multiple detections must be carefully considered. We wish to avoid having the sample dominated by a small number of sources with multiple detections of the same outflow at different velocities. For example, the UFO in PDS~456 appears at slightly different velocities (from $\sim0.23$--0.33$c$) in almost every observation taken of the source \citep{Gofford14}, most likely due to a flux-dependent velocity shift \citep{Matzeu17}.% Including every one of these measurements would strongly bias our results. 
However, an additional layer of ultra-fast absorption is present in PDS~456 \citep{Reeves18_pds456} simultaneously, which should be included separately. This is further complicated by the transient behaviour of outflowing absorption lines \citep[e.g.][]{Cappi09}, which is likely due to the gas being fully ionized at high source fluxes \citep{Pinto18} and inhomogeneities in the wind.
To mitigate this, we combine multiple UFO detections from a single source into an averaged value when the velocity estimates are within the reported errors of each other, within 0.01$c$ of each other, or within 10\% of each other, whichever is largest.
Where the same authors have written multiple papers on a particular source, taking multiple velocity measurements, we assume that the later papers supersede earlier ones, and only use the latest values unless the outflows presented are clearly distinct. Otherwise, we include all values in our analysis.

\subsection{Relativistic reflection}

Obtaining a corresponding set of reflection measurements is simpler than the UFO case, as there should only be one value of the inclination for each AGN. We take the latest available measurement for each source, unless there is an earlier \emph{NuSTAR} paper, in which case we use that value (in general, \emph{NuSTAR}'s high energy coverage gives a better constraint on the reflection spectrum than soft instruments).

\citet{Reynolds14} gives a list of `quality control' criteria for AGN spin measurements based on reflection spectroscopy. In brief, \citeauthor{Reynolds14} suggests that: a full ionized reflection model must be used; the iron abundance must be a free parameter; the inclination parameter must be free, and constrained; and the emissivity index must be free to vary and physically plausible.
For the sake of keeping as large a sample as possible, we do not implement these as a strict requirement. Rather, we select all sources with a constrained inclination and flag those which do not meet these criteria. Then, in Section~\ref{sec_results}, we run the analysis on the full sample and on those results that meet the quality control criteria.

Where the errors on the inclination are less than 5\textdegree, we make the conservative assumption that the true uncertainty is dominated by systematic errors of $\pm5$\textdegree \citep[see e.g. the difference between \textsc{relxill} and \textsc{reflionx} discussed in][]{Middleton16}. This applies in most cases. Finally, we exclude sources that require retrograde spin or significant disk truncation. In this case it is likely that the line is dominated by a narrow core, so the parameters are not reliable.
The observed inclinations are strongly concentrated at 40--50\textdegree, which is most likely due to selection effects. At higher inclinations, the LOS is likely to intersect the torus, obscuring the nucleus, and at lower inclinations the relativistic blurring is weaker and correspondingly harder to measure.

\begin{table*}
\caption{Outflow velocities and inclinations for the sources in our sample. Velocities with multiple references are the result of taking the weighted average of multiple measurements that meet the criteria discussed in Section~\ref{sec_sample_ufos}. Similarly, sources with multiple velocities are those where multiple measurements did not meet the criteria for merging.} 
\label{tab_data}
\begin{tabular}{l l l l l}
\hline
Name	&	 $v_\mathrm{UFO}$ (c)	&	Reference 	&	 $i$ (degrees) 	&	 Reference\\
\hline
1H 0419-577	&	$0.079\pm0.007$	&	\citet{Tombesi11}	&	$49.0^{+7.0}_{-4.0}$	&	\citet{Walton13_2}\\
1H 0707-495$^\ddagger$	&	$0.11^{+0.01}_{-0.02}$	&	\citet{Dauser12}&	$43.0\pm2.0$	&	\citet{Kara15}\\
	&	$0.18\pm0.01$	&	\citet{Hagino16}; \citet{Dauser12}\\
3C 111$^\ddagger$	&	$0.105\pm0.006$	&	\citet{Gofford13}; \citet{Tombesi11_3c111}	&	$44.0\pm2.0$	&	\citet{Tombesi11_3c111}$^\dagger$\\
% 3C 390.3	&	$0.145\pm0.007$	&	G13	&	$35.0\pm0$	&	W13/G01\\ This inclination is from the jet
Ark 120	&	$0.29\pm0.02$	&	\citet{Tombesi11}	&	$45.0^{+5.0}_{-2.0}$	&	\citet{Garcia14}\\
IC 4329A	&	$0.098\pm0.004$	&	\citet{Tombesi11}	&	$35.0\pm5.0$	&	\citet{Mantovani14}$^\dagger$\\
IRAS 00521-7054$^\ddagger$ &	$0.403^{+0.007}_{-0.006}$	&	Walton et al. (in prep)	&	$63^{+3}_{-2}$	&	Walton et al. (in prep)\\
IRAS 13224-3809$^\ddagger$	&	$0.236\pm0.006$	&	\citet{Parker17_nature}	&	$59.0\pm1.0$	&	\citet{Parker17_nature}\\
IRAS 13349+2438$^\ddagger$	&	$0.13\pm0.01$	&	Parker et al. (submitted)	&	$48.0^{+2.0}_{-1.0}$	&	Parker et al. (submitted)\\
MCG-5-23-16	&	$0.116\pm0.004$	&	\citet{Tombesi11}	&	$51.0\pm7.0$	&	\citet{Zoghbi17}\\
MR 2251-178	&	$0.137\pm0.008$	&	\citet{Gofford13}&	$24.0^{+3.0}_{-5.0}$	&	\citet{Nardini14}\\
Mrk 1044$^\ddagger$	&	$0.10\pm0.01$	&	Mallick et al. (submitted)	&	$47\pm3$	& Mallick et al. (submitted)\\
Mrk 509	&	$0.14\pm0.0024$	&	\citet{Cappi09}; \citet{Tombesi11}	&	$50.0^{+5.0}_{-3.0}$	&	\citet{Walton13_2}\\
	&	$0.171\pm0.003$	&	\citet{Cappi09}; \citet{Tombesi11}\\
	&	$0.197\pm0.005$	&	\citet{Cappi09}; \citet{Tombesi11}\\
Mrk 766	&	$0.039\pm0.03$	&	\citet{Gofford13}	&	$39.0^{+6.0}_{-3.0}$	&	Buisson et al. (submitted)\\
	&	$0.082\pm0.006$	&	\citet{Tombesi11}\\
Mrk 79	&	$0.092\pm0.004$	&	\citet{Tombesi11}	&	$24.0\pm1.0$	&	\citet{Gallo11_mrk79}\\
Mrk 841	&	$0.055\pm0.025$	&	\citet{Tombesi11}	&	$46.0^{+6.0}_{-5.0}$	&	\citet{Walton13_2}\\
NGC 4051	&	$0.202\pm0.006$	&	\citet{Tombesi11}	&	$37.0\pm5.0$	&	Risaliti et al. (in prep)\\
NGC 4151	&	$0.0452\pm0.0099$	&	\citet{Gofford13}; \citet{Patrick12}	&	$<10$	&	\citet{Beuchert17}\\
	&	$0.106\pm0.007$	&	\citet{Tombesi11}\\
NGC 5506	&	$0.246\pm0.006$	&	\citet{Gofford13}	&	$41.0^{+0.1}_{-0.2}$	&	\citet{Sun17}\\
PDS 456$^\ddagger$	&	$0.278\pm0.003$	&	\citet{Reeves18_pds456}; \citet{Matzeu17}	&	$65.0\pm2.0$	&	\citet{Chiang17}\\
	&	$0.46\pm0.02$	&	\citet{Reeves18_pds456}\\
PG 1211+143$^\ddagger$	&	$0.0598\pm0.00069$	&	\citet{Pounds16}; \citet{Danehkar18}; 	&	$44.0\pm2.0$	&	\citet{Lobban16}\\
&	&	\citet{Reeves18_pg1211}\\
	&	$0.129\pm0.002$	&	\citet{Pounds16}\\
	&	$0.151\pm0.003$	&	\citet{Tombesi11}\\
Swift J2127	&	$0.231\pm0.006$	&	\citet{Gofford13}	&	$49.0\pm2.0$	&	\citet{Marinucci14_swiftj2127}\\
\hline
\end{tabular}

% B17: \citet{Beuchert17}, 
% B$^{*}$: Buisson et al. (submitted), 
% C09: \citet{Cappi09}, 
% C17: \citet{Chiang17}, 
% D12: \citet{Dauser12}, 
% D18: \citet{Danehkar18}, 
% G11: \citet{Gallo11_mrk79}, 
% G13: \citet{Gofford13}, 
% G14: \citet{Garcia14}, 
% H16: \citet{Hagino16}, 
% K15: \citet{Kara15}, 
% L16: \citet{Lobban16}, 
% Mn14: \citet{Mantovani14}, 
% Mr14: \citet{Marinucci14_swiftj2127},
% M17: \citet{Matzeu17}, 
% N14: \citet{Nardini14}, 
% P12: \citet{Patrick12}, 
% P16: \citet{Pounds16}, 
% P17: \citet{Parker17_nature}, 
% P$^{*}$: Parker et al. (submitted), 
% R18a: \citet{Reeves18_pg1211}, 
% R18b: \citet{Reeves18_pds456}, 
% R$^*$: Risaliti (in prep.), 
% S17: \citet{Sun17}, 
% T11a: \citet{Tombesi11_3c111}, 
% T11b: \citet{Tombesi11}, 
% W13: \citet{Walton13_2},
% W$^*$: Walton (in prep),
% Z17: \citet{Zoghbi17}.\\
$^\dagger$These results do not meet the quality-control criteria of \citet{Reynolds14}.\\
$^\ddagger$These sources have joint reflection/UFO fitting, which either gives the result reported here or is consistent with it.
\end{table*}

\section{Results}
\label{sec_results}
\subsection{Correlation analysis}

% In Fig.~\ref{fig_correlation} we show the UFO velocity against reflection inclination, along with the best-fit linear relation, calculated using Bayesian regression to take into account the upper limits on $i$. A positive trend is evident, but strongly reliant on a single point at high velocity \citep[the second UFO zone in PDS~456, ][]{Reeves18_pds456}.

% \begin{figure}
% \centering
% \includegraphics[width=\linewidth]{UFO_relation.pdf}
% \caption{Words.}
% \label{fig_correlation}
% \end{figure}
We show the UFO velocity against reflection inclination points in Figs~\ref{fig_model} and~\ref{fig_diskmodel}, overplotted with simple models (see Section~\ref{sec_models}).
We use a Monte-Carlo approach to estimate the significance of any correlation between the two parameters. The distributions of $v$ and $i$ are approximately log-normal and normal, respectively, so we randomly draw 100,000 sets of points from distributions with the same mean and standard deviation (Mean log($v$)$=-0.89$, $\sigma_{\log(v)}=0.29$, mean $i=44.4$, $\sigma_i=9.53$). Of these simulated sets of points, we find 37 that exceed the Pearson correlation coefficient of the real data (0.64) and 629 that exceed the Spearman correlation coefficient (0.52). This gives probabilities of 0.0004 and 0.006 for a correlation this strong occurring from randomly distributed points. Excluding the two sources where the reflection modeling does not meet the \citet{Reynolds14} criteria marginally strengthens the correlation (Pearson $r=$0.63, $P=0.0011$, Spearman $r=0.49$, $P=0.017$). 
% Cutting this part. Dom told me about his new UFO, so PDS456 is not the only high velocity one any more.
%However, if we remove PDS~456, which has the highest inclination and by far the highest velocity in our sample, the correlation is significantly weaker. We find Pearson and Spearman correlation coefficients of 0.31 and 0.29, and corresponding chance probabilities of 0.136 and 0.158.

\subsection{Models}
\label{sec_models}

\begin{figure}
\centering
\includegraphics[width=0.9\linewidth]{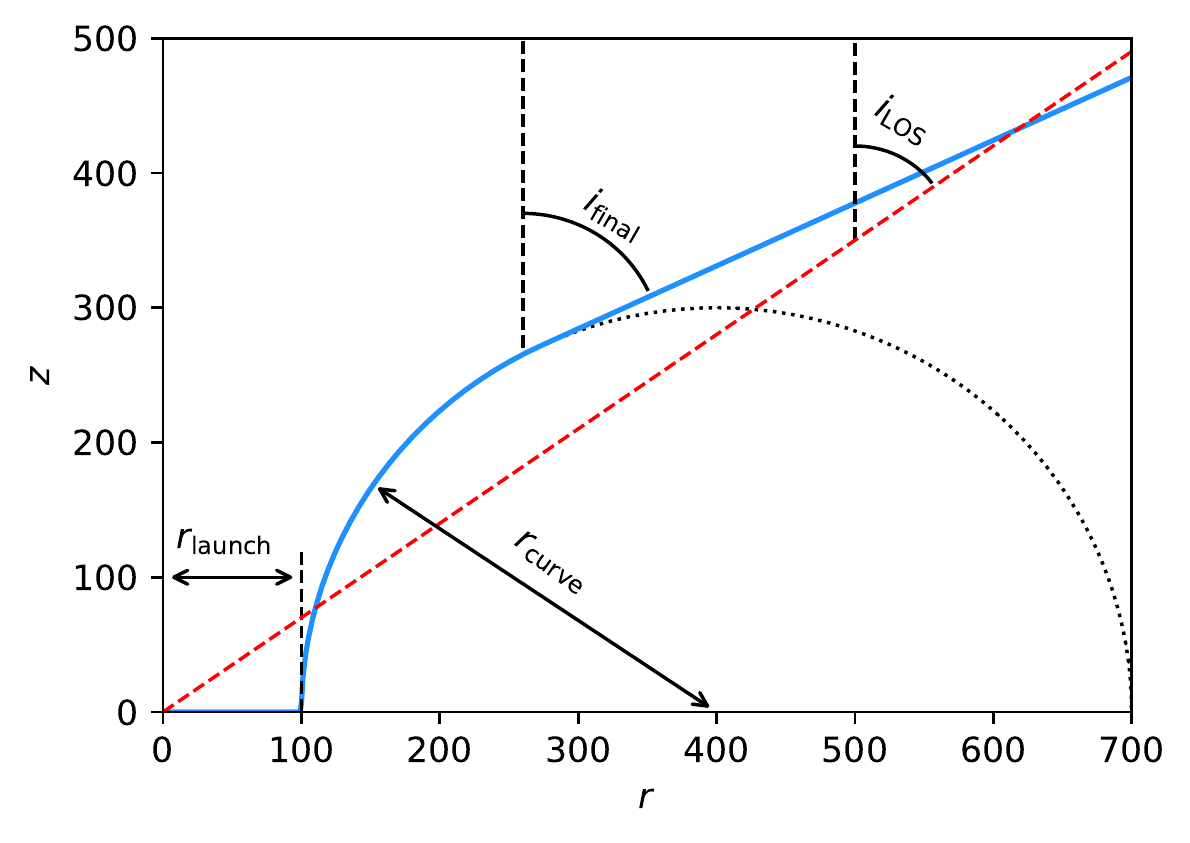}
\caption{Simple streamline geometry, with the wind shown in blue and the line of sight in red (dashed). Length units are arbitrary.}
\label{fig_geometry}
\end{figure}

Regardless of whether a linear correlation is a significant improvement over the null hypothesis of randomly distributed points, it is still possible to use these points to infer something about the geometry of UFOs. %In this section, we construct toy models of outflow geometries and examine the angular dependence of the observed velocity.
We construct a simple imitation of a stream-line, where a thin outflow starts moving vertically from the disc at some radius $r_\mathrm{launch}$, following a circular path with radius $r_\mathrm{curve}$. For this purpose, the units of radius are arbitrary. Once it reaches a final inclination $i_\mathrm{final}$ it leaves the circular path and travels on the tangent with $i=i_\mathrm{final}$. We also assume that the UFO accelerates along this path, following $v= v_\mathrm{inf}(1-R_v/(R_v+r))^\beta$ \citep[adapted from][]{Knigge95,Sim08}, where $\beta$ is a constant (and $\beta=0$ gives a constant velocity) and $R_v$ is a characteristic length scale. We assume that the X-ray source is coincident with the black hole ($h=0$, $r=0$), and ignore all relativistic effects. This geometry is shown in Fig.~\ref{fig_geometry}. This geometry is intended as an approximation of that in radiation-driven winds \citep[e.g.][]{Proga00}, as these give a simple explanation for the higher velocities at higher inclinations. MHD winds predict concave stream lines \citep{Fukumura10}, so give higher velocities at small inclinations. However, the exact pattern observed depends on the ionization and density structure within the wind, so it may still be possible to explain these results in an MHD scenario.

\begin{figure*}
\includegraphics[width=0.85\linewidth]{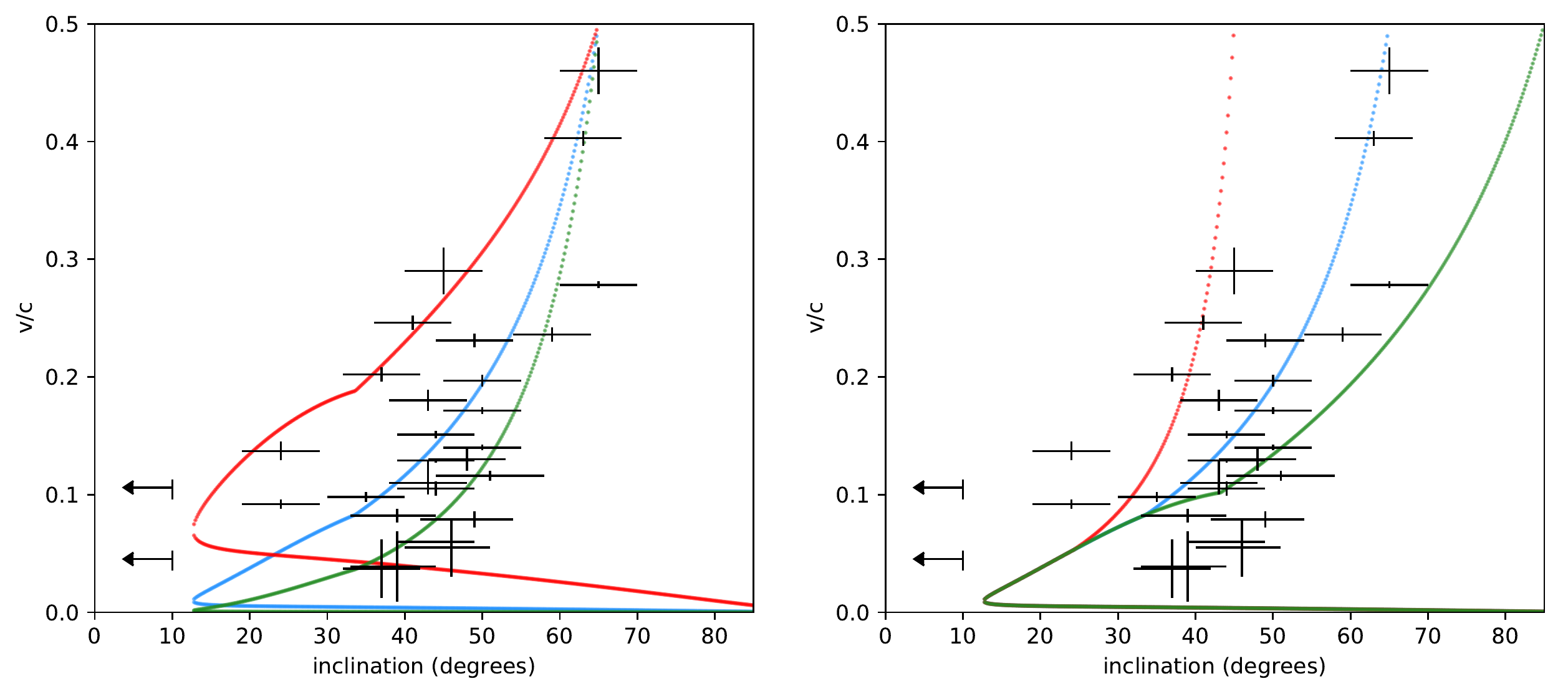}
\caption{UFO velocity as a function of reflection inclination, overplotted with a toy model for an outflowing disc wind. Left: the effect of changing the acceleration coefficient $\beta$ to 0.5 (red,left), 1.0 (blue, middle) and 1.5 (green, right) with final inclination $i_\mathrm{final}=65^\circ$ . Right: the effect of setting $i_\mathrm{final}$ to 45$^\circ$ (red, left), 65$^\circ$ (blue, middle) and 85$^\circ$ (green, right) with $\beta=1$.}
\label{fig_model}
\end{figure*}

In most cases where the line of sight intersects the wind in this model, it crosses the wind twice. Once while the wind is rising steeply, and once in the tail where the gradient is constant. Because of the radial acceleration assumed and close alignment with the line of sight in the tail, this intersection results in a much higher apparent velocity. A simple way of only producing one measurable value for the velocity is to assume that the gas where the first intersection with the LOS occurs is fully ionized. In this case, no absorption lines would be produced, and only the second intersection would be observed.
An example that provides a reasonable match to the data is shown in Fig.~\ref{fig_model}, with parameters $r_\mathrm{launch}=10$, $r_\mathrm{curve}=300$, $v_\mathrm{inf}=0.5c$, and $R_v=1000$. We show the effect of varying the acceleration coefficient $\beta$ and final inclination $i_\mathrm{final}$. From this it is clear that meaningful constraints on these parameters can be obtained.

An alternative way of giving a relation between the velocity of the absorption line and the viewing angle is to produce the absorption in the disc itself. This model, explored by \citet{Gallo11_ufos, Gallo13_ufos}, explains UFO absorption using a surface layer on the disc, with the strong blueshift due to the orbital velocity of the absorbing material, rather than an outflowing wind. In this case, the inclination dependence of the absorption velocity arises from the increased LOS velocity of the disc at high inclinations. 
\begin{figure}
\centering
\includegraphics[width=0.9\linewidth]{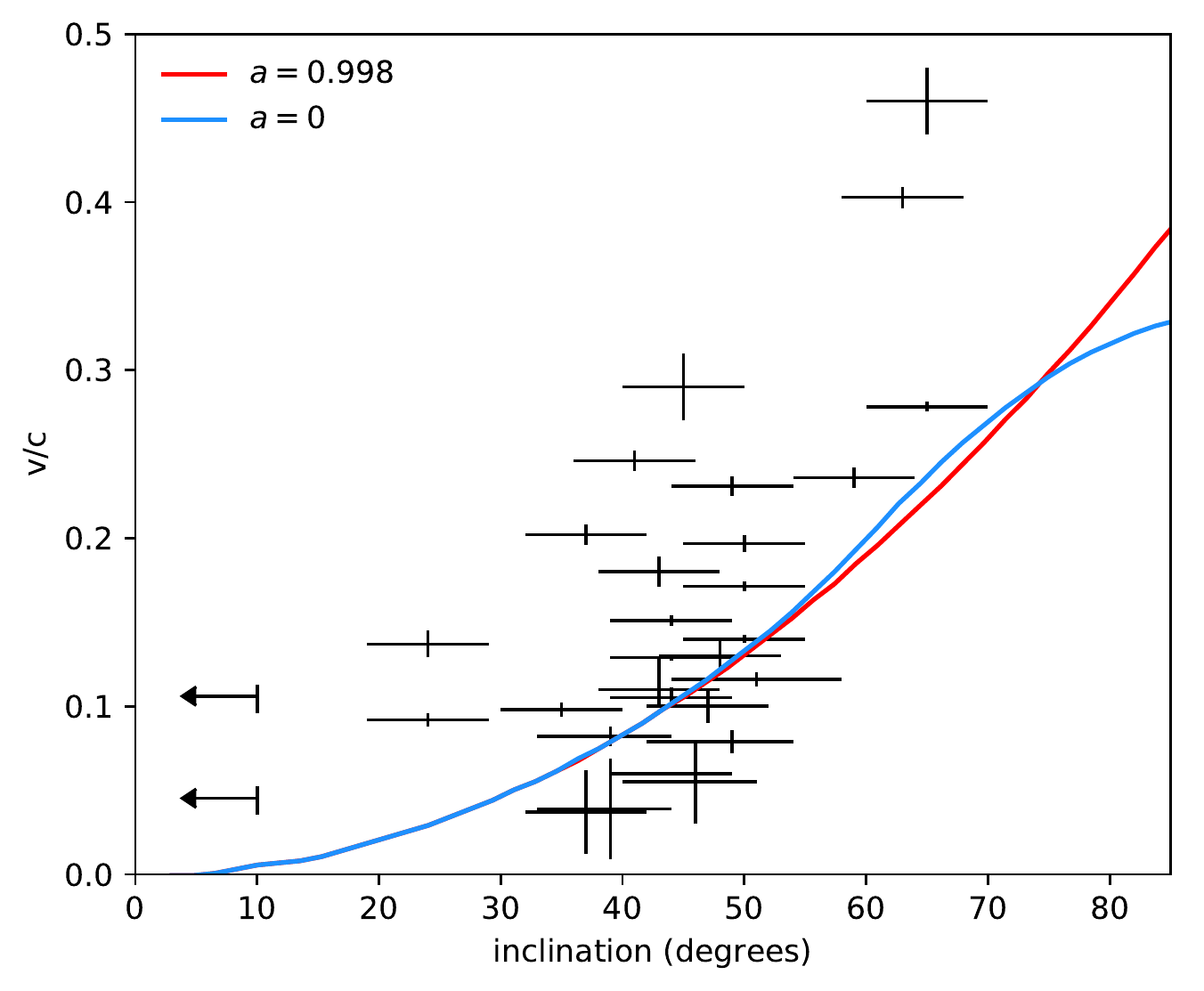}
\caption{Toy model for absorption from a layer on the disc.}
\label{fig_diskmodel}
\end{figure}
As a simple proxy for the velocity of an absorption line from the surface of the disc, we take the maximum blueshift from the \textsc{relline} model \citep{Dauser10}. This gives a simple correlation between $i$ and $v$ (shown in Fig.~\ref{fig_diskmodel}), although it does not reach high enough velocities to account for the most blueshifted absorption lines. The only parameter of this model is the black hole spin ($a$), but this has a limited effect as the maximum blueshift of the disc comes from further out than the innermost stable circular orbit. We note that this predicted velocity should be an upper limit as it is the maximum found on the disc, so points should generally lie below the line.

\section{Discussion}

% Cutting this, as above.
%The significance of the possible correlation between inner-disc inclination and UFO velocity depends strongly on the highest velocity, highest inclination point, from PDS~456 \citep{Reeves18_pds456}. There is no consensus on the origin of the broad Fe emission feature in this source, however. 
We note that there are other possible explanations for broad emission lines in AGN. For example, \citet{Nardini15} interpret the emission line in PDS~456 as a P-Cygni profile, where the Fe~K emission is produced by scattering off the outflowing wind. In this case, the relativistic broadening is produced by the velocity of the wind instead of the orbits in the disc \citep[][]{Done07_pcyg}. This model relies on partial-covering absorption to explain most of the spectral complexity. However, \citet{Chiang17} present an alternative model where the broad-band spectrum can be fully explained by relativistic reflection, warm absorption, and the UFO. This interpretation is supported by the detection of a soft X-ray lag, generally interpreted as reverberation in the inner disc and found in many unobscured sources \citep{DeMarco13}.
While the detection of an X-ray lag is usually considered strong evidence for the presence of relativistic reflection in a source \citep[and many of the sources in our sample show reverberation lags,][]{Kara16, DeMarco13}, it is difficult to rule out a contribution to the total broad-line profile from scattering in the wind. Indeed, given that disc winds are likely to be most dense at the point they launch from the disc and should be co-rotating with it, then scattering from the disc surface and wind may be thought of as a single continuous process. The impact of having Fe~K emission from both the disc and wind is not understood, but could have a significant effect on the measured inclination. A higher velocity outflow will produce more blueshifted emission, in the same way that a higher inclination gives a more blueshifted line profile for disc reflection. %Without publicly available models, this is hard to explore further, but 
We note that the tentative relation identified here may be indicative of the effect of wind emission on the net relativistic line profile rather than an inclination dependence.
A related issue is the lack of simultaneous UFO/reflection modeling. Using a simplified phenomenological continuum may exaggerate the significance of line features, leading to false detections \citep{Zoghbi15}. Similarly, not accounting for UFO absorption during reflection modeling may bias the measured parameters. We will revisit the spectra of some of these sources for joint modeling to investigate this further in future work.

We have assumed throughout that all UFOs have the same shape and velocity profile, and that the observed velocity is solely determined by the LOS angle. This is unlikely to be the case in practice: the accretion rate, for example, is likely to have a major impact on the velocity of the material. Similarly, we have implicitly assumed that wind instabilities play a negligible role in determining the observed velocity, which is unlikely.
Another caveat is that our sample is anything but unbiased, and the biases are poorly understood. Reflection measurements are generally biased towards high spin \citep{Vasudevan16}, and there may be a similar bias towards high inclination, as it produces broader, easier to measure lines. Similarly, it is plausible that UFO velocity measurements are biased towards lower velocities, as the sensitivity of X-ray detectors typically declines with energy (although this may be remedied as the \emph{NuSTAR} archive grows).

\section{Conclusions}
We have identified a correlation between the velocity of highly ionized absorption features from UFOs and the inclination of the inner accretion disc measured from reflection spectroscopy. The correlation is formally significant, but heavily reliant on a small number of points at high velocity.

We show that the observed points can be explained by simple toy models of an outflowing wind or absorption from a disc, although the latter cannot account for the highest velocity features. With more detailed modeling and higher quality data, this technique could be very powerful for constraining the geometry of outflowing material in AGN.

\section*{Acknowledgements}
MLP and GM are supported by European Space Agency (ESA) Research Fellowships. JJ acknowledges support by the Cambridge Trust and the Chinese Scholarship Council Joint Scholarship Programme (201604100032). DJKB is supported by an STFC studentship.

%%%%%%%%%%%%%%%%%%%%%%%%%%%%%%%%%%%%%%%%%%%%%%%%%%



\bibliographystyle{mnras}
%\bibliography{bibliography_ufoinclinations} % if your bibtex file is called example.bib

% \begin{thebibliography}{99}
% \bibitem[\protect\citeauthoryear{Author}{2012}]{Author2012}
% Author A.~N., 2013, Journal of Improbable Astronomy, 1, 1
% \bibitem[\protect\citeauthoryear{Others}{2013}]{Others2013}
% Others S., 2012, Journal of Interesting Stuff, 17, 198
% \end{thebibliography}

%%%%%%%%%%%%%%%%%%%%%%%%%%%%%%%%%%%%%%%%%%%%%%%%%%

%%%%%%%%%%%%%%%%% APPENDICES %%%%%%%%%%%%%%%%%%%%%

% \appendix

% \section{Some extra material}

% If you want to present additional material which would interrupt the flow of the main paper,
% it can be placed in an Appendix which appears after the list of references.

%%%%%%%%%%%%%%%%%%%%%%%%%%%%%%%%%%%%%%%%%%%%%%%%%%

% Don't change these lines
\bsp	% typesetting comment
\label{lastpage}
\end{document}